\RequirePackage{fix-cm}
\documentclass{svjour3}                     
\smartqed  
\usepackage{graphicx}
\newcommand{\nn}{\nonumber}
\newcommand{\be}{\begin{equation}}
\newcommand{\ee}{\end{equation}}
\newcommand{\bea}{\begin{eqnarray}}
\newcommand{\eea}{\end{eqnarray}}

\newcommand{\ep}{\epsilon}

\newcommand{\om}{\omega}

\newcommand{\ov}{\overline}

\newcommand{\vk}{\vec k}

\newcommand{\vq}{\vec q} 
\newcommand{\vl}{\vec l}

\newcommand{\mn}{\mu\nu}
\newcommand{\del}{\partial}

\newcommand{\unit}{1\!\!1}
\begin{document}

\title{Shear viscosity of pionic and nucleonic
components from their different possible mesonic and baryonic thermal fluctuations 
\thanks{This work is financed by Funda\c{c}\~ao de Amparo \`a Pesquisa do Estado de 
S\~ao Paulo - FAPESP, Grant Nos. 2012/16766-0.}
}

\titlerunning{Shear viscosity from mesonic and baryonic fluctuations}        

\author{Sabyasachi Ghosh       
}


\institute{Sabyasachi Ghosh \at
            Instituto de F\'{\i}sica Te\'orica, Universidade Estadual Paulista,
            Rua Dr. Bento Teobaldo Ferraz, 271 - Bloco II, 01140-070 S\~ao Paulo, 
            SP, Brazil   \\
              \email{sabyaphy@gmail.com}           
}

\date{Received: date / Accepted: date}

\maketitle

\begin{abstract}
Owing to the Kubo relation, the shear viscosities of pionic and 
nucleonic components have been evaluated from their corresponding 
retarded correlators of viscous stress tensor in the static limit,
which become non-divergent only for the non-zero thermal widths of the 
constituent particles. In the real-time thermal field theory,
the pion and nucleon thermal widths have respectively been obtained from
the pion self-energy for different meson, baryon loops and
the nucleon self-energy for different pion-baryon loops.
We have found a non-monotonic momentum distributions of pion and nucleon
thermal widths, which have been integrated out by their respective Bose-enhanced
and Pauli-blocked phase space factors during evaluation of their shear viscosities.
The viscosity to entropy density ratio for this mixed gas of pion-nucleon system
decreases and approaches toward its lower bound as the temperature and
baryon chemical potential increase within the relevant domain of hadronic matter.
\end{abstract}
%
%
%
%
%
\section{Introduction}
\label{sec:intro}
A strongly interacting matter is expected to be produced, 
instead of a weakly interacting gas, 
at RHIC and LHC energies as the shear viscosity 
of the matter thus produced is exposed to be very small.
This was concluded by the hydrodynamical 
simulations~\cite{Romatschke1,Heinz1,Roy,Niemi,LHC_Hydro1,LHC_Hydro2} 
as well as some transport calculations~\cite{Xu1,Greco1,Greco2,LHC_Transport} 
to explain the elliptic flow parameter observed at RHIC and LHC.
According to the investigations conducted in
Refs.~\cite{Hufner,Csernai,Gyulassy,Kapusta:2008vb,Chen1,Purnendu,Chen2}, 
the shear viscosity to entropy density ratio, $\eta/s$ may reach a minimum 
in the vicinity of a phase transition, which is also indicated by some
lattice QCD calculations\cite{Lat1,Lat2,Lat3}. 
The minimum value of $\eta/s$
may be very close to its quantum lower 
bound, commonly known as the KSS bound~\cite{KSS}.
Owing to these interesting issues, a growing interest in the 
microscopic calculation of shear viscosity for the 
QGP phase~\cite{Arnold,QGP2,Moore,Basagoiti,Arts,SG_NJL} and hadronic 
phase~\cite{Dobado1,Dobado2,Muronga,Nakano,Nicola,Itakura,Gorenstein,Greiner,SPal,Toneev,Prakash_2012,Buballa,Weise,SSS,Krein,Prakash_2013,Denicol,Bass,HM}
has been noticed in recent times,
though the transport coefficient calculations of 
nuclear matter started somewhat 
earlier~\cite{Gavin,Prakash,Toneev_7,Toneev_8,Toneev_9,Toneev_10}. 
Importance of knowing the explicit temperature dependence
of shear viscosity for hadronic phase has been pointed out in
a recent work by Niemi et 
al.~\cite{Niemi}. They have shown that the extracted 
transverse momentum $p_T$ dependence of elliptic 
flow parameter, $v_2(p_T)$, of RHIC data is highly sensitive to the temperature 
dependent $\eta/s$ in hadronic matter and almost independent of the viscosity in 
the QGP phase. 

Being inspired by this, we have studied 
the shear viscosity of pionic medium~\cite{GKS} and 
then extended our study to the pion-nucleon 
system~\cite{G_N}. To calculate shear viscosity of 
the pionic component via Kubo relation~\cite{Zubarev,Kubo}, 
the thermal correlator of viscous stress tensor for pionic constituents
has to be derived, where a finite thermal width of pion
should be included for getting a non-divergent value of correlator
in the static limit~\cite{Nicola,Weise,G_IJMPA}. In Ref.~\cite{GKS}, the pion
thermal width is estimated from the pion self-energy
for different mesonic loops, which are obtained in the formalism of
real-time thermal field theory (RTF). Similarly, the thermal correlator 
of viscous stress tensor for nucleonic constituents is obtained in
Ref.~\cite{G_N} to calculate the shear viscosity of nucleonic component,
where different pion-baryon loops are taken into account to determine the nucleon
thermal width. Now, in the two component nucleon-pion system,
pion propagation may also have some baryonic fluctuations besides
the mesonic fluctuations. This contribution, which was absent in our
previous studies~\cite{GKS,G_N}, is considered in the present work
to revisit our shear viscosity calculation for two component nucleon-pion system.

In the next section, the formalism
of shear viscosity for pionic and nucleonic components are
briefly described, where their corresponding thermal widths
are discussed in three different subsections. In the subsection~\ref{subsec:pi_bar},
we have elaborately deduced the pion thermal width in baryonic medium
by calculating the pion self-energy for different baryonic loops
in the formalism of RTF. In the next two subsections, the relevant
expressions of pion thermal width due to different mesonic loops
and nucleon thermal width from different pion-baryon loops are
briefly addressed as the detailed deduction of these expressions
are already provided in the previous studies~\cite{GKS,G_N}.
The numerical outcomes are discussed in Sec.~\ref{sec:num}
and in Sec.~\ref{sec:concl}, we have summarized and concluded this article.
\section{Formalism}
\label{sec:form}
Owing to the famous Kubo formula~\cite{Zubarev,Kubo},
the spectral function of two point viscous-stress tensor, 
$\pi^{\mn}$ can determine
the shear viscosity in momentum
space by the standard relation~\cite{Nicola} 
\be
\eta=\frac{1}{20}\lim_{q_0,\vq \rightarrow 0}\frac{1}{q_0}
\int d^4x e^{iq\cdot x}\langle[\pi_{ij}(x),\pi^{ij}(0)]\rangle_\beta~,
\label{eta_Nicola}
\ee
where $\langle \hat{O}\rangle_\beta$ for any operator $\hat{O}$
denotes the equilibrium ensemble average;
$\langle \hat{O}\rangle_\beta={\rm Tr}\frac{e^{-\beta H}\hat{O}}{{\rm Tr}e^{-\beta H}}$.

\begin{figure}
\begin{center}
\includegraphics[scale=0.52]{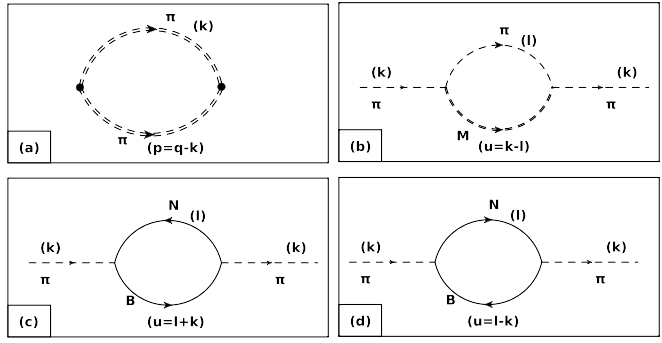}
\caption{The diagram (a) is a schematic one-loop representation of viscous-stress
tensor for the medium with pionic constituents. The double dashed lines
for the pion propagators indicate that they have some finite thermal
width, which can be derived from the pion self-energy diagrams (b), (c) and (d).
The diagram (b) represents pion self-energy for mesonic ($\pi M$) loops.
Direct and cross diagrams of pion self-energy 
for $NB$ loops are represented by (c) and (d) respectively.} 
\label{eta_pi_N}
\end{center}
\end{figure}

\begin{figure}
\begin{center}
\includegraphics[scale=0.52]{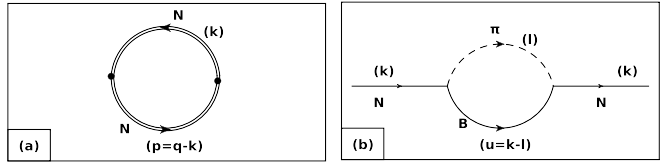}
\caption{The diagram (a) is a schematic one-loop representation of viscous-stress
tensor for the medium with nucleonic constituents and
the diagram (b) represents nucleon self-energy for $\pi B$ loops.} 
\label{eta_N_piB}
\end{center}
\end{figure}

The simplest one-loop expressions of Eq.~(\ref{eta_Nicola}) 
for pion and nucleon degrees of freedom
are respectively given below~\cite{G_IJMPA}
\be
\eta_\pi=\frac{\beta I_\pi}{30\pi^2}\int^{\infty}_{0} 
\frac{d\vk\vk^6}{{\om^\pi_k}^2\Gamma_\pi}n_k(\om^\pi_k)
\{1+n_k(\om_k^\pi)\} 
\label{eta_pi}
\ee
and
\bea
\eta_N&=&\frac{\beta I_N}{15\pi^2}\int^{\infty}_{0} 
\frac{d\vk\vk^6}{{\om^N_k}^2\Gamma_N}[n^+_k(\om_k^N)\{1-n^+_k(\om^N_k)\}
+n^-_k(\om^N_k)\{1-n^-_k(\om^N_k)\}]~.
\nn\\
\label{eta_N}
\eea
Their schematic diagrams are shown in Fig.~\ref{eta_pi_N}(a) and \ref{eta_N_piB}(a)
respectively.
Hence, adding the pionic and nucleonic components, we get the total shear viscosity
\be
\eta_{\rm T}=\eta_{\pi}+\eta_N~.
\ee
In the above equations, $n_k(\om^\pi_k)=1/\{e^{\beta\om^\pi_k}-1\}$ is 
Bose-Einstein (BE) distribution of pion with energy 
$\om^\pi_k=(\vk^2+m_\pi^2)^{1/2}$ whereas 
$n^{\pm}_k=1/\{e^{\beta(\om^\pi_k\mp \mu_N)}+1\}$ are Fermi-Dirac (FD) distributions
of nucleon and anti-nucleon with energy $\om^N_k=(\vk^2+m_N^2)^{1/2}$. 
The corresponding thermal widths, $\Gamma_\pi$ and $\Gamma_N$ 
for pion and nucleon can be defined as
\bea
\Gamma_\pi&=&\sum_B\Gamma_{\pi(NB)} + \sum_M\Gamma_{\pi(\pi M)}
\nn\\
&=&-\sum_B{\rm Im}{\Pi}^R_{\pi(NB)}(k_0=\om^\pi_k,\vk)/m_\pi
-\sum_M{\rm Im}{\Pi}^R_{\pi(\pi M)}(k_0=\om^\pi_k,\vk)/m_\pi
\nn\\
\label{Gam_pi}
\eea
and
\be
\Gamma_N=\sum_B\Gamma_{N(\pi B)}=-\sum_B{\rm Im}\Sigma^R_{N(\pi B)}(k_0=\om^N_k,\vk)
\label{Gam_N}
\ee
respectively, where ${\Pi}^R_{\pi(NB)}(k)$ is pion self-energy for different 
nucleon-baryon ($NB$) loops (shown in Fig.~\ref{eta_pi_N}(c) and (d)),
${\Pi}^R_{\pi(\pi M)}(k)$ is pion self-energy for different pion-meson ($\pi M$)
loops (shown in Fig.~\ref{eta_pi_N}(b)) 
and $\Sigma^R_{N(\pi B)}(k)$ is nucleon self-energy for different 
pion-baryon ($\pi B$) loops (shown in Fig.~\ref{eta_N_piB}(b)).
The superscript $R$ stands for retarded component of self-energy
and subscripts represent the external (outside the bracket) and 
internal (inside the bracket) particles for the corresponding
self-energy graphs as shown in Fig.~\ref{eta_pi_N}(b), (c), (d)
and Fig.~\ref{eta_N_piB}(b).
Adoption of finite thermal widths $\Gamma_\pi$ and $\Gamma_N$ in Eq.~(\ref{eta_pi})
and (\ref{eta_N}) respectively
is a very well established technique~\cite{Nicola,Weise},
which is generally used in Kubo approach
to get a non-divergent value of the shear viscosity coefficient.
In this respect this treatment is equivalent to quasi-particle approximation
or relaxation time approximation.
Again, this one-loop expression of $\eta_\pi$ or $\eta_N$
from Kubo approach~\cite{Nicola,Weise,S_rev} exactly coincides with the
expression coming from the relaxation-time approximation of
the kinetic theory approach~\cite{Gavin,Prakash,SSS,S_rev}.
Hence, the thermal width of medium constituent plays a vital
role in determining the  numerical strength of shear viscosity of the medium. 

%
Next we discuss the calculations of thermal widths from different one-loop self-energy
graphs as shown in Fig.~(\ref{eta_pi_N}) and (\ref{eta_N_piB}).

\subsection{Pion thermal width for different baryonic loops}
\label{subsec:pi_bar}
Let us first concentrate on the pion
self-energy calculations for different possible baryon loops {\it i.e} ${\Pi}^R_{\pi(NB)}$.
During propagation in the medium,
pion propagator can undergo different intermediate NB loops, 
where $B=\Delta(1232)$, $N^*(1440)$, $N^*(1520)$,
$N^*(1535)$, $\Delta^*(1600)$, $\Delta^*(1620)$, $N^*(1650)$, 
$\Delta^*(1700)$, $N^*(1700)$, $N^*(1710)$, $N^*(1720)$ are
accounted in this work.
The masses of all the 4-star baryon resonances (in MeV) are presented inside the brackets.
The direct and cross diagrams of pion self-energy 
for $NB$ loops have been represented
in the diagrams~\ref{eta_pi_N}(c) and (d).

In real-time formalism of thermal field theory (RTF), self-energy becomes
$2\times 2$ matrix with $11$, $12$, $21$ and $22$ components. From any of
the components, one can found the retarded part of self-energy, which is
directly related with physical quantity - thermal width (inverse of thermal
relaxation time). 
Let us start with the $11$-component of in-medium pion 
self-energy for $NB$ loop :
\be
\Pi^{11}_{\pi(NB)}(k)=i \sum_{a=-1,+1}
\int \frac{d^4l}{(2\pi)^4}  L(k,l)
E^{11}_N(l) E^{11}_B(l-a k) 
\label{Pi11B}
\ee
where $E^{11}_N(l)$ and $E^{11}_B(l-ak)$ are
scalar parts of the nucleon and 
baryon propagators respectively at finite temperature. 
In RTF this expression is as follows
\be
E^{11}_N(l)=\frac{-1}{l^2-m_N^2+i\eta}-2\pi i\{n^+_l\theta(l_0)
+n^-_l\theta(-l_0)\}\delta(l^2-m_N^2)~,
\label{de11}
\ee
where $n^{\pm}_l(\om^N_l)=1/\{e^{\beta(\om^N_l \mp \mu_N)}+1\}$
are the FD distributions of nucleon and 
anti-nucleon for energy $\om^N_l=(\vl^2+m_N^2)^{1/2}$
and $\mu_N$ is the chemical potential of nucleon which is 
supposed to be equal with the chemical potentials
of all the baryons considered here. 
The two values of $a$ in Eq.~(\ref{Pi11B}) correspond to the direct 
and crossed diagrams, shown in Fig.~\ref{eta_pi_N}
(c) and (d) respectively, which can be obtained from 
one another by changing the 
sign of external momentum $k$.

Let us first discuss diagram (d) for which $a=+1$.
Integrating Eq.~(\ref{Pi11B}) over $l^0$
and using the relation,
\be
{\rm Im}\Pi^{R}_{\pi(NB)}(k)={\rm tanh}(\beta k_0/2){\rm Im}\Pi^{11}_{\pi(NB)}(k)~,
\ee
the retarded component of the in-medium self energy (imaginary part) 
can be expressed as
\bea
{\rm Im}{\Pi}^R_{\pi(NB)}(k)&=&\pi\ep(k_0)\int\frac{d^3l}{(2\pi)^3}
\frac{1}{4\om^N_l\om^B_u}
\nn\\
&&L_1[\{1-n_l^+(\om^N_l)-n_u^-(\om^B_u)\}\delta(k_0 -\om^N_l-\om^B_u)
\nn\\
&&~~+\{n_l^+(\om_l^N)-n_u^+(\om^B_u)\}\delta(k_0-\om^N_l+\om^B_u)] 
\nn\\
&&~~+ L_2[\{-n_l^-(\om^N_l) +n_u^-(\om^B_u)\}\delta(k_0 +\om^N_l-\om^B_u)
\nn\\
&&+\{-1+n_l^-(\om^N_l)+n_u^+(\om^B_u)\}\delta(k_0 +\om^N_l+\om^B_u)~,
\label{Pi_a}
\eea
where $n^{\pm}_u(\om^B_u)=1/\{e^{\beta(\om^B_u \mp \mu_N)}+1\}$ 
are also FD distribution functions
for baryon and anti-baryon with $\om^B_u=\{(\vl-\vk)^2+m_{B}^2\}^{1/2}$
and $L_{1,2}$ denote the values of $L(l_0,\vl,k)$ for
$l_0=\om^N_l$ and $-\om^N_l$ respectively. 
The different $\delta$ functions in Eq.~(\ref{Pi_a}) create the regions of 
different branch cuts in $k_0$-axis viz.
$-\infty$ to $-\{\vk^2+(m_N+m_B)^2\}^{1/2}$ for unitary cut in negative $k_0$-axis,
$-\{\vk^2+(m_B-m_N)^2\}^{1/2}$ to $\{\vk^2+(m_B-m_N)^2\}^{1/2}$ for Landau cut and 
$\{\vk^2+(m_N+m_B)^2\}^{1/2}$ to $\infty$ for unitary cut in positive $k_0$-axis.
In these different kinematic regions,
the imaginary part of the pion self-energy becomes non-zero.
Among the four terms in the right hand side of Eq.~(\ref{Pi_a}), the third term
contributes in pion thermal width for baryonic loops, 
$\Gamma_{\pi(NB)}$ because the pion pole ($k_0=\om^\pi_k,\vk$) is
situated within the Landau cut $(0$ to $\{\vk^2+(m_B-m_N)^2\}^{1/2}~)$ 
in the positive $k_0$-axis.
From the Eq.~(\ref{Gam_pi}), using the relation,
\be
\Gamma_{\pi(NB)}=-{\rm Im}{\Pi}^R_{\pi(NB)}(k_0=\om^\pi_k,\vk)/m_\pi
\ee
and adding the relevant Landau cut contributions of both diagrams (c) 
and (d), the total thermal width of pion for any $NB$ loop is given by
\bea
\Gamma_{\pi(NB)}(\vk,T,\mu_N) &=& \frac{1}{16\pi|\vk| m_\pi} \int^{\om^N_{l-}}_{\om^N_{l+}} 
d\om^N_l\times 
L\left(l_0=-\om^N_l,\vl,
k_0=\om_k^\pi,\vk
\right)[\{-n^+_l(\om^N_l) 
\nn\\
&&+ n^+_u(\om^B_u=\om^\pi_k + \om^N_l)\}
+\{-n^-_l(\om^N_l) 
+ n^-_u(\om^B_u=\om^\pi_k + \om^N_l)\}]~,
\nn\\
\label{G_pi_NB}
\eea
where 
\be
\om^N_{l\pm} = \frac{S^2_{\pi(NB)}}{2m_\pi^2} 
\left(- \om^\pi_k \pm |\vk| \, W_{\pi(NB)} \right) ,
\ee
with 
\be
S^2_{\pi(NB)}=m_\pi^2-m_B^2+m_N^2
\ee
and 
\be
W_{\pi(NB)} = \left(1- {4m_\pi^2m_N^2}/{S^4_{\pi(NB)}}\right)^{1/2}~.
\ee
Lagrangian densities of spin $J_B=1/2$ and $3/2$ baryons can be written as
\be
{\cal L}_{\rm Baryon}=\sum_B\left[{\cal L}^{free}_{B(J_B=1/2,3/2)}
+ {\cal L}^{int}_{B(J_B=1/2,3/2)}\right]~,
\ee
where free parts of Lagrangian densities for baryonic fields with spin 
$J_B=1/2$ and $J_B=3/2$ are
\bea
{\cal L}^{\rm free}_{B(J_B=1/2)}&=&\sum_{B(J_B=1/2)}{\ov \psi_B}(i\gamma^\mu\del_\mu-m_B)\psi_B~,
\nn\\
{\cal L}^{\rm free}_{B(J_B=3/2)}&=&\sum_{B(J_B=3/2)}-\frac{1}{2}{\ov \psi^\mu_B}
(\epsilon_{\mu\nu\alpha\beta}\gamma^\alpha\del^\beta-im_B\sigma^{\alpha\beta})\psi^\nu_B
\nn\\
&&{\rm with}~\sigma^{\alpha\beta}=\frac{i}{2}\left[\gamma^\alpha,\gamma^\beta\right]
\eea
and their interaction parts are~\cite{Leupold},
\bea
{\cal L}^{\rm int}&=&\frac{f}{m_\pi}{\ov \psi}_B\gamma^\mu
\left\{
\begin{array}{c}
i\gamma^5 \\
\unit
\end{array}
\right\}
\psi_N\del_\mu\pi + {\rm h.c.}~{\rm for}~J_B^P=\frac{1}{2}^{\pm},
\nn\\
{\cal L}^{\rm int}&=&\frac{f}{m_\pi}{\ov \psi}^\mu_B
\left\{
\begin{array}{c}
\unit \\
i\gamma^5
\end{array}
\right\}
\psi_N\del_\mu\pi + {\rm h.c.}~{\rm for}~J_B^P=\frac{3}{2}^{\pm},
\label{Lag_BNpi}
\eea
Here $P$ stands for parity quantum numbers of the baryons. The coupling constants
$\pi NB$ interactions are fixed from the experimental decay widths of 
$B\rightarrow N\pi$ channels~\cite{G_pi_JPG}. They are $f/m_\pi=15.7$, $2.5$,
$11.6$, $1.14$, $3.4$, $1.22$, $1.14$, $9.5$, $2.8$, $0.35$ and $1.18$
for $\Delta(1232)$, $N^*(1440)$, $N^*(1520)$,
$N^*(1535)$, $\Delta^*(1600)$, $\Delta^*(1620)$, $N^*(1650)$, 
$\Delta^*(1700)$, $N^*(1700)$, $N^*(1710)$, $N^*(1720)$.
Using (\ref{Lag_BNpi}), we have found the vertex factors~\cite{G_pi_JPG}:
\bea
L(k,l)
&=&-4\left(\frac{f}{m_\pi}\right)^2[2(k\cdot l)^2-a(k\cdot l)k^2
-k^2(l^2+m_Nm_B)],~~ {\rm for}~ J_B^P=\frac{1}{2}^{\pm},
\nn\\
&=&-\frac{8}{3m_B^2}\left(\frac{f}{m_\pi}\right)^2[m_Nm_B+l^2-a(k\cdot l)]
[(l\cdot k-ak^2)^2-k^2m_B^2],
\nn\\
&&~~~~~~~~~~~~~~~~~~~~~~~~~~~~~~~~~~~~~~~~~~
~~~~~~~~~~~~~~~~~~~~~{\rm for} J_B^P=\frac{3}{2}^{\pm}~.
\eea

\subsection{Pion thermal width for different mesonic loops}
\label{subsec:pi_mes}
To calculate the mesonic loop contribution of pionic thermal width
$\Gamma_{\pi(\pi M)}$, we have
evaluated pion self-energy for $\pi M$ loops, where $M$ stands for $\sigma$
and $\rho$ mesons. This contribution estimated in our previous work~\cite{GKS}
elaborately. Following that~\cite{GKS}, the expression of pion thermal
width from the mesonic loops is given below
\bea
\Gamma_{\pi(\pi M)}(\vk,T) &=& \frac{1}{16\pi|\vk| m_\pi} \int^{\om^\pi_{l-}}_{\om^\pi_{l+}} 
d\om^\pi_l  L\left(l_0=-\om^\pi_l,\vl,
k_0=\om_k^\pi,\vk
\right)\{n_l(\om^\pi_l) 
\nn\\
&&~~~~~~~~~~~~~~~~~~~~~~~~~~~~~
~~~~- n_u(\om^M_u=\om^\pi_k + \om^\pi_l)\}~,
\label{G_pi_piM}
\eea
where $n_l$, $n_u$ are BE distribution functions of $\pi$, $M$
mesons respectively and the limits of integration are
\be
\om^\pi_{l\pm} = \frac{S^2_{\pi(\pi M)}}{2m_\pi^2} 
\left(- \om^\pi_k \pm |\vk| \, W_{\pi(\pi M)} \right) ,
\ee
with 
\be
S^2_{\pi(\pi M)}=2m_\pi^2-m_M^2
\ee
and 
\be
W_{\pi(\pi M)} = \left(1- {4m_\pi^4}/{S^4_{\pi(\pi M)}}\right)^{1/2}~.
\ee
Lagrangian density of pion, sigma and rho mesons can be written as
\be
{\cal L}={\cal L}^{\rm free}_\pi + {\cal L}^{\rm free}_\sigma + {\cal L}^{\rm free}_\rho 
+  {\cal L}^{\rm int}_{\pi\pi\rho} + {\cal L}^{\rm int}_{\pi\pi\sigma}~,
\ee
where free parts of Lagrangian densities for pseudo-scalar ${\vec \pi}$, scalar $\sigma$ 
and vector $\rho^\mu$ fields are
are 
\bea
{\cal L}^{\rm free}_\pi&=&\frac{1}{2}\{(\del_\mu{\vec \pi})\cdot(\del^\mu{\vec \pi})-m_\pi^2{\vec \pi}^2\}
\nn\\
{\cal L}^{\rm free}_\sigma&=&\frac{1}{2}\{(\del\sigma)^2-m_\sigma^2\sigma^2\}
\nn\\
{\cal L}^{\rm free}_\rho&=&\frac{1}{2}\{(\rho_{\mn}\rho^{\mn})-m_\rho^2(\rho_\mu\rho^\mu)\},~
\rho^{\mn}=(\del^\mu\rho^\nu-\del^\nu\rho^\mu)
\eea
and their interaction parts are~\cite{Weise_Lag,GKS,SSS}
\bea
{\cal L}^{\rm int}_{\pi\pi\rho} &=& g_\rho \, {\vec \rho}_\mu \cdot {\vec \pi} \times \del^\mu {\vec \pi} 
\nn\\
{\cal L}^{\rm int}_{\pi\pi\sigma}&=& \frac{g_\sigma}{2} m_\sigma {\vec \pi}\cdot {\vec\pi}\,\sigma~.
\label{Lag_pipiM}
\eea
The coupling constant $g_\rho=6$ and $g_\sigma=5.82$ are fixed from experimental
decay width~\cite{GKS} and physical masses of pion, sigma and rho mesons are taken
as $m_\pi=0.140$ GeV, $m_\sigma=0.390$ GeV and $m_\rho=0.770$ GeV.
Using (\ref{Lag_pipiM})
we have obtained the vertex factors:
\bea
L(k,l) &=& - \frac{g^2_\sigma m_\sigma^2}{4}, 
~~~~~~~~~~~~~~~~~~~{\rm for}~M=\sigma~,
\nn\\
 &=& -\frac{g^2_\rho}{m_\rho^2} \, 
[ k^2 \left(k^2 - m^2_\rho\right) + 
l^2 \left(l^2 - m^2_\rho\right) 
- \, 2\{ (k\cdot l) \, m^2_\rho + k^2 \,l^2 \}],~{\rm for}~M=\rho~.
\nn\\
\eea

\subsection{Nucleon thermal width}
\label{subsec:N}
In order to calculate the nucleonic thermal width
$\Gamma_{N(\pi B)}$, we have
evaluated nucleon self-energy for different possible $\pi B$ loops, 
where $B$ stands for all the baryons as taken in pion self-energy
for baryonic loops.
This contribution is rigorously addressed in our previous work~\cite{G_N}.
Hence taking the relevant expression of nucleon thermal 
width for any $\pi B$ loop from the Ref.~\cite{G_N}, we have
\bea
\Gamma_{N(\pi B)}(\vk,T,\mu_N) &=& \frac{1}{16\pi|\vk| m_\pi} \int^{\om^\pi_{l-}}_{\om^\pi_{l+}} 
d\om^\pi_l L\left(l_0=-\om^\pi_l,\vl,
k_0=\om_k^N,\vk
\right)\{n_l(\om^\pi_l) 
\nn\\
&&~~~~~~~~~~~~~~~~~~~~~~
~~~~~~~~~+ n_u(\om^B_u=\om^N_k + \om^\pi_l)\}~,
\label{G_N_piB}
\eea
where $n_l$ and $n_u$ are BE and FD distribution functions for $\pi$ and $B$ respectively.
The relevant limits of integration in Eq.~(\ref{G_N_piB}) are:
\be
\om^N_{l\pm} = \frac{S^2_{N(\pi B)}}{2m_N^2} 
\left(- \om^N_k \pm |\vk| \, W_{N(\pi B)} \right) ,
\ee
with 
\be
S^2_{N(\pi B)}=m_N^2-m_B^2+m_\pi^2
\ee
and 
\be
W_{N(\pi B)} = \left(1- {4m_N^2m_\pi^2}/{S^4_{N(\pi B)}}\right)^{1/2}~.
\ee
The vertex factors~\cite{G_N}:
\bea
L(k,l)&=&-\left(\frac{f}{m_\pi}\right)^2\left\{\left(\frac{R^2}{2}-m_\pi^2
\right)l_0 -Pm_\pi^2m_B\right\}
~{\rm for}~J_B^P=\frac{1}{2}^{\pm}~,
\nn\\
L(k,l)&=&-\left(\frac{f}{m_\pi}\right)^2\frac{2}{3m_B^2}
\left\{\left(\frac{R^2}{2}-m_\pi^2\right)^2
-m_\pi^2m_B^2\right\}(k_0-l_0+Pm_B)
~{\rm for}~J_B^P=\frac{3}{2}^{\pm}
\nn\\
\eea
can be deduced by using the $\pi NB$ interaction Lagrangian 
densities from Eq.~(\ref{Lag_BNpi}).
%
%
%
\section{Results and Discussion}
\label{sec:num}
The detailed Landau cut contributions of pion
self-energy for mesonic loops and nucleon self-energy
for different $\pi B$ loops are investigated in the earlier
Refs.~\cite{GKS} and \cite{G_N}, where their corresponding
contributions in the shear viscosity are also addressed.
Now in the two component pion-nucleon system, another
contribution to pion thermal width can arise from
the pion self-energy with baryonic loops, which was not 
considered in our previous studies of the shear viscosity~\cite{GKS,G_N}.
The main purpose of the present work is to
include these baryonic loop contributions in the pion thermal width and
to revisit the shear viscosity results.

\begin{figure}
\begin{center}
\includegraphics[scale=0.35]{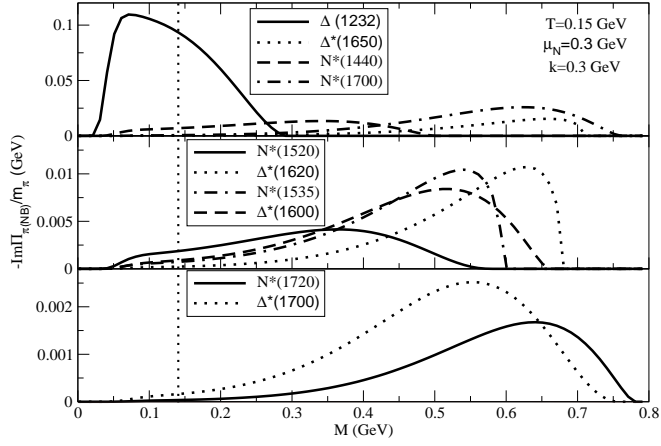}
\caption{$\Gamma_{\pi(NB)}(M_k)$ for different $NB$ loops
in their Landau regions, which contain the pion pole $M_k=m_\pi$,
denoted by straight dotted line.} 
\label{self_M}
\end{center}
\end{figure}
\begin{figure}
\begin{center}
\includegraphics[scale=0.35]{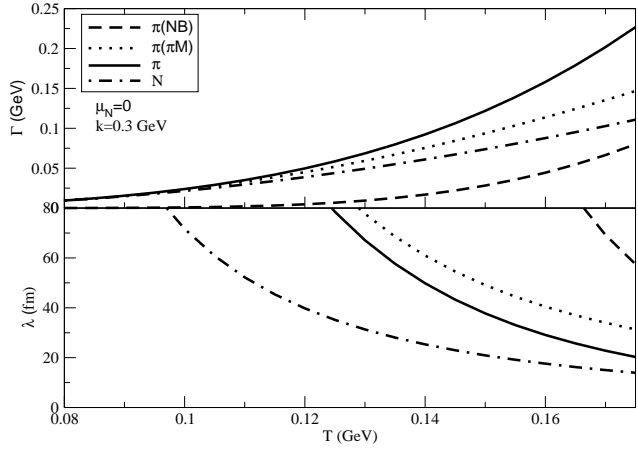}
\caption{The on-shell thermal widths (upper panel) and mean free
paths (lower panel) of pion for $\pi M$ loops, $NB$ loops and their total
are represented by dotted, dashed and solid lines respectively while 
dash-dotted line denotes the same results for nucleon component with
all possible $\pi B$ loops.} 
\label{gm_T_piN}
\end{center}
\end{figure}
\begin{figure}
\begin{center}
\includegraphics[scale=0.35]{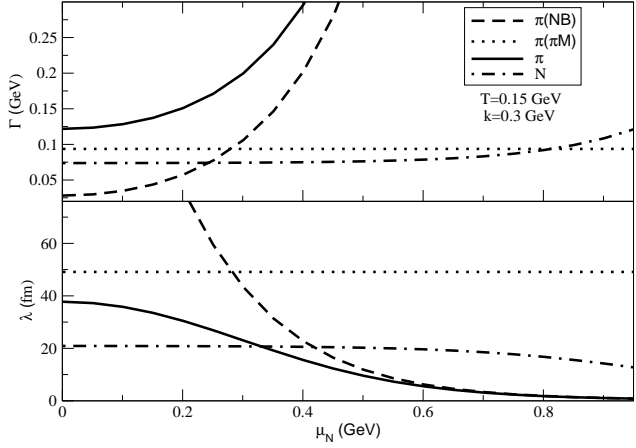}
\caption{Same as Fig.~(\ref{gm_T_piN}) along the $\mu_N$ axis.} 
\label{gm_mu_piN}
\end{center}
\end{figure}
\begin{figure}
\begin{center}
\includegraphics[scale=0.35]{gm_k_piN.eps}
\caption{Same as Fig.~(\ref{gm_T_piN}) along the $\vk$ axis} 
\label{gm_k_piN}
\end{center}
\end{figure}
Let us first zoom in our attention on the $\Gamma_{\pi(NB)}$.
Fig.~(\ref{self_M}) represents the Landau cut contributions of
different $NB$ loops on the invariant mass axis $M_k$, which can be numerically
generated by replacing $\om^\pi_k=(\vk^2+M_k^2)^{1/2}$ in Eq.~(\ref{G_pi_NB}).
For a fixed set of parameters $\vk$, $T$ and $\mu_N$,
the $\Gamma_{\pi(NB)}(M_k)$ for baryons $B=\Delta(1232)$, $\Delta^*(1650)$,
$N^*(1440)$, $N^*(1700)$ in the upper panel, 
$B=N^*(1520)$, $\Delta^*(1620)$, $N^*(1535)$, $\Delta^*(1600)$ 
in the middle panel and $B=N^*(1720)$, $\Delta^*(1700)$
in the lower panel are individually presented in the Fig.~(\ref{self_M}).
The Landau cut regions, where $\Gamma_{\pi(NB)}(M_k)$ for different $NB$ 
loops have attained their non-zero values, are clearly observed
in the invariant mass axis. As an example, for $N\Delta$ loop 
the Landau region is from $M=0$ to $(m_\Delta-m_N)=0.292$ GeV.
The straight dotted line denotes position of pion pole (i.e. $M_k=m_\pi$),
which indicates the on-shell contribution of $\Gamma_{\pi(NB)}$ for 
different baryonic loops. Here we identify $N\Delta$ loop as a leading
candidate to contribute in the pion thermal width among all the baryonic loops.

Adding the on-shell contribution of all $NB$ loops, we get the 
total thermal width of pion for baryonic loops, which is plotted 
against temperature by dashed line in the upper panel of 
Fig.~(\ref{gm_T_piN}). Using Eqs.(\ref{G_pi_piM}) and
(\ref{G_N_piB}), we have also generated the numerical values of
total $\Gamma_{\pi(\pi M)}(T)$ (dotted line) and 
$\Gamma_{N(\pi B)}(T)$ (dashed line) by
adding their corresponding loop contributions for a same set
of input parameters ($\vk=0.3$ GeV, $\mu_N=0$). Following 
Eq.~(\ref{Gam_pi}), the solid line represents the $T$ dependence of 
total thermal width of pion, $\Gamma_\pi(T)$ after adding the 
mesonic and baryonic loop contributions. 
All of the $\Gamma$'s are monotonically increasing function but with different
rate of increment.
The corresponding results of 
mean free path, defined by $\lambda=\vk/(\om_k\Gamma)$, are
presented in the lower panel of the Fig.~(\ref{gm_T_piN}).
Being inversely proportional to the thermal width, the mean free paths for all of the 
components monotonically decrease with $T$ and exhibit divergent nature
at low $T$. 
Along the $\mu_N$ axis, $\lambda$'s ($\Gamma$'s) for all of the components
also decrease (increase) with different rates as shown in the lower (upper)
panel of Fig.~(\ref{gm_mu_piN}). Here we see that independent nature
of pion thermal width ($\Gamma_{\pi(\pi M)}$) or mean free path ($\lambda_{\pi(\pi M)}$) 
for mesonic loops is transformed
to an increasing or decreasing nature when the baryonic loop contribution
is added.
Moreover, the divergence problem of $\lambda_{\pi(NB)}(\mu_N)$ at low $\mu_N$
is also cured in the total mean free path for pionic component $\lambda_\pi(\mu_N)$.
A mild $\mu_N$ dependence of the nucleonic component is observed.

At fixed values of $T$ and $\mu_N$, the momentum distribution of 
thermal widths (upper panel) and mean free paths (lower panel) for 
all of the components have been displayed in Fig.~(\ref{gm_k_piN}). 
Being equivalent to the momentum distribution for the imaginary part 
of optical potential (see e.g.~\cite{G_pi_JPG,Rapp_pi}), thermal width
of pion for any mesonic or baryonic loop exhibits a non-monotonic
distribution with a peak structure along the $\vk$ axis. The mathematical
reason can roughly be understood from the relevant 
Eqs.~(\ref{G_pi_piM}) and (\ref{G_pi_NB}) as described in Ref.~\cite{G_pi_JPG}.
After adding the different $NB$ loop contributions, each of which has similar
kind of momentum distribution with different numerical strength, we get
a multi-peak complex structure of $\Gamma_{\pi(NB)}(\vk)$.
When we add it with the $\Gamma_{\pi(\pi M)}(\vk)$, which
contains a dominating profile with one peak (due to $\pi\rho$ loop mainly),
then a well behaving momentum distribution with less complex structures (solid line)
is obtained. The $\Gamma_N(\vk)$ (dash-dotted line) approximately appears
constant with a mild reduction with $\vk$. 
Though we notice a divergent nature of $\lambda_{\pi(NB)}(\vk)$
out side the range of $\vk=0.1-1$ GeV
but the total $\lambda_\pi$ in the entire momentum range
remains non-divergent or finite with an well-behaved distribution.

%
\begin{figure}
\begin{center}
\includegraphics[scale=0.35]{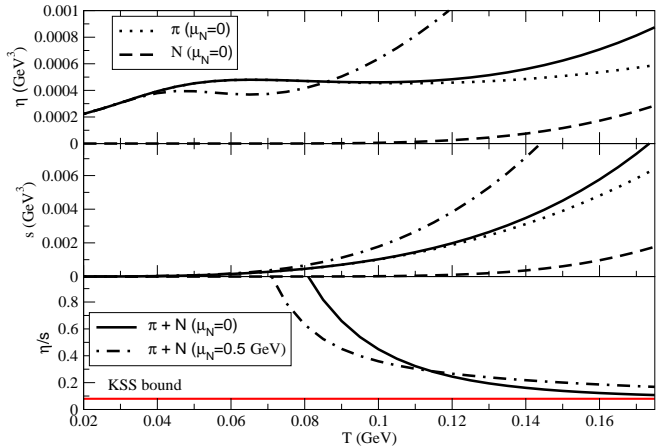}
\caption{Temperature dependence of shear viscosities (upper panel)
and entropy densities (middle panel)
for pionic (dotted line), nucleonic (dashed line) components and
their total at two different nucleon chemical potentials:
$\mu_N=0$ (solid line) and $\mu_N=0.5$ GeV (dash-dotted line).
In the lower panel, the ratios of total viscosity to entropy density
are represented as a function of $T$
at same set of $\mu_N$'s, taken in the upper and middle panels.} 
\label{eta_s_T}
\end{center}
\end{figure}
\begin{figure}
\begin{center}
\includegraphics[scale=0.35]{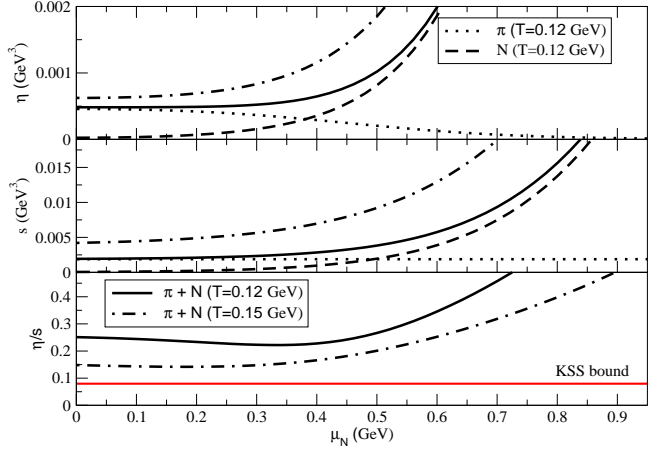}
\caption{The $\mu_N$ dependence of same quantities as Fig.~(\ref{eta_s_T})
at two different temperatures: $T=0.12$ GeV (solid line)
and $T=0.15$ GeV (dash-dotted).} 
\label{eta_s_mu}
\end{center}
\end{figure}
\begin{figure}
\begin{center}
\includegraphics[scale=0.35]{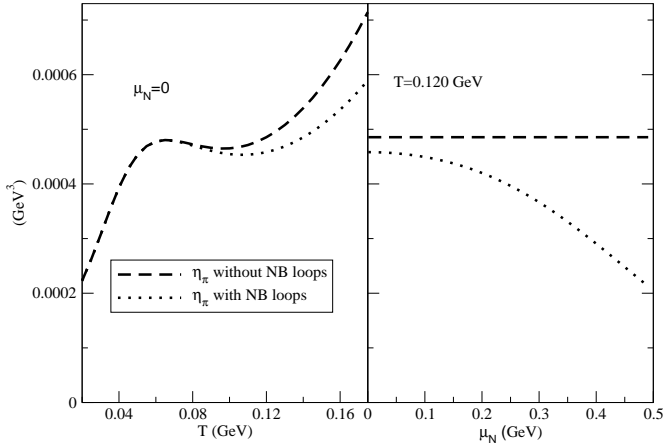}
\caption{Dashed and dotten lines show $\eta_\pi$ without and
with baryonic fluctuations in pion propagation respectively.} 
\label{ref1}
\end{center}
\end{figure}
Using the total thermal width for pionic component, $\Gamma_\pi(\vk,T,\mu_N)$
in Eq.~(\ref{eta_pi}) and
for nucleonic component, $\Gamma_{N}(\vk,T,\mu_N)$ in Eq.~(\ref{eta_N}),
we have obtained shear viscosities $\eta_\pi(T,\mu_N)$ and $\eta_N(T,\mu_N)$.
They are plotted by dotted and dashed lines respectively as functions
of $T$ and $\mu_N$ in the upper panels of Fig.~(\ref{eta_s_T}) and (\ref{eta_s_mu}).
After exhibiting a soft peak structure in the low $T(<0.1$ GeV), $\eta_\pi(T)$
monotonically increases with a very mild rate in the high $T(>0.1$ GeV) domain.
When a monotonically increasing function $\eta_N(T)$
is added with this pionic component, the total shear viscosity 
$\eta_T$ in high $T$ domain 
enhances with slightly larger rate of increment, which can be noticed 
by solid line in the upper panel of Fig.~(\ref{eta_s_T}).
Another curve of $\eta_T(T)$ at $\mu_N=0.5$ GeV is shown by dash-dotted
line, which faces a rapid increment after $T=0.06$ GeV. The reason of this
drastic enhancement can be well understood from the $\mu_N$ dependence
of the two components $\eta_\pi(\mu_N)$ and $\eta_N(\mu_N)$. The upper
panel of Fig.~(\ref{eta_s_mu}) exposes a rapidly increasing function
$\eta_N(\mu_N)$ and a soft decreasing function $\eta_\pi(\mu_N)$.
Remembering the Fig.~(\ref{gm_mu_piN}), we can understand that the origin
of soft decreasing nature of $\eta_\pi(\mu_N)$ is coming from the baryonic
loop contribution of pion as its mesonic loop contribution is independent
of $\mu_N$. Right panel of Fig.~(\ref{ref1}) is zooming this fact more distinctly,
where we see how inclusion of NB loops in pion self-energy makes $\eta_\pi$
deviate from its independent (dashed line) to dependent (dotted line) nature
with $\mu_N$. This is the main and dramatically important contributions of 
the present article as an extension of earlier works~\cite{GKS,G_N}. After
including it, providing a complete picture of shear viscosity calculation
for pion-nucleon system is the main aim of this present investigation.

The dotted lines in the left and right panel of Fig.~(\ref{ref1}) are
exactly same as dotted lines in the upper panels of Fig.~(\ref{eta_s_T})
and (\ref{eta_s_mu}) respectively. Still those curves are repeated for 
elaborating the effect of baryonic fluctuations in pion self-energy.
As the phase space factor of Eq.~(\ref{eta_pi}) does not depend
on the $\mu_N$, so only thermal width $\Gamma_\pi(\mu_N)$ controls on the 
$\mu_N$ dependence of $\eta_\pi$. Now between two components $\Gamma_{\pi(\pi M)}$
and $\Gamma_{\pi(NB)}$ of $\Gamma_\pi$, the latter one has only the dependency
of $\mu_N$ as exposed in Fig.~(\ref{gm_mu_piN}).
The decreasing nature of $\eta_\pi(\mu_N)$
is solely governed by the increasing (decreasing) nature of function $\Gamma_\pi(\mu_N)$
($\lambda_\pi(\mu_N)$).
Whereas, in case of Eq.~(\ref{eta_N}), nucleonic phase space factor depend
on $\mu_N$ so strongly that it makes $\eta_N(\mu_N)$ be an increasing function
after dominating over the opposite action of $\Gamma_N(\mu_N)$ or 
$\lambda_N(\mu_N)$ on the $\eta_N(\mu_N)$.

Middle panels of Fig.~(\ref{eta_s_T}) and (\ref{eta_s_mu}) represent
the $T$ and $\mu_N$ dependence of entropy densities for pionic and 
nucleonic components by following their ideal expressions:  
\be
s_\pi = 3\beta\int \frac{d^3\vk}{(2\pi)^3} \left(\om^\pi_k+\frac{\vk^2}{3\om^\pi_k}\right)
n_k(\om^\pi_k)
\label{s_pi}
\ee
and
\be
s_N=4\beta\int\frac{d^3\vk }{(2\pi)^3}
\left(\om^N_k+\frac{\vk^2}{3\om^N_k}-\mu_N\right)n^+_k(\om^N_k)~.
\label{s_N}
\ee
Using these numerical results of entropy densities (middle panel) 
as well as for shear viscosities (upper panel) for
pionic, nucleonic components
and their total , we have presented their 
corresponding ratios in the lower panels of the graphs, where straight
horizontal (red) lines stand for KSS bound of the ratio.
The decreasing nature of ratio is sustained for both $\mu_N=0$ (solid line)
and $\mu_N=0.5$ GeV (dash-dotted line) in the entire $T$ axis. 
The former is dominating over the later in magnitude for $T\leq 0.12$ GeV
and then an opposite trend is followed beyond $T=0.12$ GeV. 
Therefore, the ratio in the $\mu_N$ axis at $T=0.15$ GeV (dash-dotted line)
and $T=0.12$ GeV (solid line) are exhibiting a nature opposite
to each other (up to $\mu_N\approx 0.4$ GeV), which can be observed in the
lower panel of Fig.~(\ref{eta_s_mu}). Nevertheless, both of them increase
in high $\mu_N$ domain ($\mu_N>0.4$ GeV). 
Most of the earlier work~\cite{Itakura,Denicol,Bass}
showed a reducing nature of ratio along the $\mu_N$ axis, which is
also found in the present work up to $T\approx 0.12$ GeV but 
beyond $T=0.120$ it is not found. It indicates that our approach
has some deficiency with respect to the earlier work~\cite{Itakura,Denicol,Bass}.
This deficiency may be the mixing effect of two component system~\cite{Itakura}, 
which have been taken care in our further investigations and discussed in next paragraph.

\begin{figure}
\begin{center}
\includegraphics[scale=0.35]{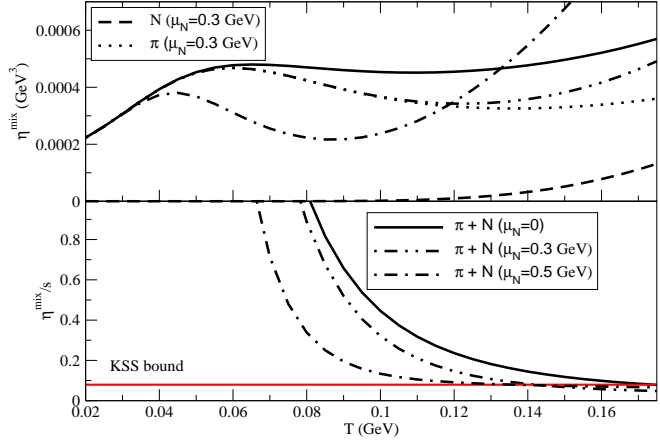}
\caption{Upper panel: Temperature dependence of shear viscosities of 
pionic (dotted line), nucleonic (dashed line) components and their
total (at three different values of $\mu_N$) in presence of mixing effect.
Lower panel: The ratios of total viscosity to entropy density vs $T$
at $\mu_N=0$ (solid line), $0.3$ GeV (dash-double-dotted line) and $0.5$ GeV
(dash-dotted line).} 
\label{etamix_T_piN}
\end{center}
\end{figure}
\begin{figure}
\begin{center}
\includegraphics[scale=0.35]{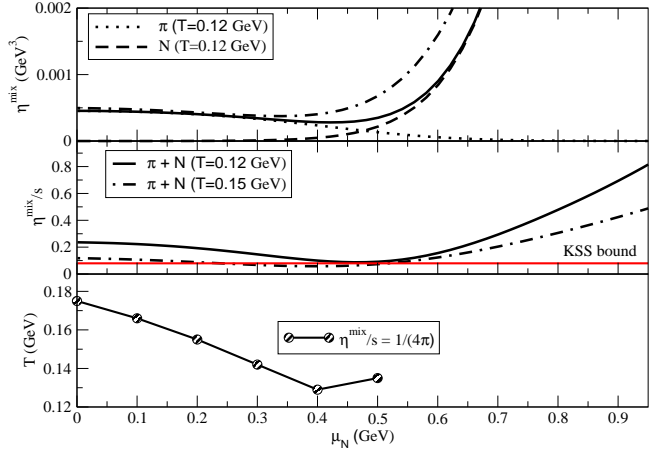}
\caption{upper and middle panels show the $\mu_N$ dependence of same quantities
as in Fig.~(\ref{etamix_T_piN}) at two different temperatures: $T=0.12$ GeV (solid line)
and $T=0.15$ GeV (dash-dotted). 
Lower panel: The different points of ($T$, $\mu_N$), where $\eta^{\rm mix}/s$
is approximately equal to the KSS bound.} 
\label{etamix_mu_piN}
\end{center}
\end{figure}

We have adopted a rough mixing effect~\cite{G_N},
which can generally be expected between two components of a mixed
gas~\cite{Itakura,mix}. From the Eqs.~(\ref{eta_pi}) and (\ref{eta_N}),
one can clearly notice that the phase space factors of 
$\eta_\pi$ and $\eta_N$ do not face any mixing effect of 
pion density, $\rho_\pi=3\int\frac{d^3k}{(2\pi)^3}n_k(\om_k^\pi)$ and
nucleon density, $\rho_N=4\int\frac{d^3k}{(2\pi)^3}n^+_k(\om_k^N)$.
Although their thermal widths $\Gamma_\pi$ and $\Gamma_N$ contain
this mixing effect as they depend on thermal distribution functions
of both, pion and nucleon. Following
the approximated relation~\cite{Itakura,G_N,mix}
\be
\eta^{\rm mix}_{\rm tot}=\eta^{\rm mix}_\pi +\eta^{\rm mix}_N~,
\label{etamix_tot}
\ee
with
\be
\eta^{\rm mix}_\pi=\frac{\eta_\pi}{1+\left(\frac{\rho_N}{\rho_\pi}\right)
\left(\frac{\sigma_{\pi N}}{\sigma_{\pi\pi}}\right)\sqrt{\frac{1+m_\pi/m_N}{2}}}
\label{etamix_pi}
\ee
and
\be
\eta^{\rm mix}_N=\frac{\eta_N}{1+\left(\frac{\rho_\pi}{\rho_N}\right)
\left(\frac{\sigma_{\pi N}}{\sigma_{NN}}\right)\sqrt{\frac{1+m_N/m_\pi}{2}}}~,
\label{etamix_N}
\ee
where the cross sections of all kind of scattering are simply taken as constant
with same order of magnitude ({\it i.e.} 
$\sigma_{\pi\pi}\approx \sigma_{\pi N} \approx \sigma_{NN}$).
In presence of this mixing scenario, the $T$ dependence
of $\eta^{\rm mix}_\pi$ (dotted line), $\eta^{\rm mix}_N$ (dashed line) and 
their total $\eta^{\rm mix}_{\rm tot}$ at $\mu_N=0$ (solid line), 
$\mu_N=0.3$ GeV (dash-dotted line) and $\mu_N=0.5$ GeV
are shown in the upper panel of Fig.~(\ref{etamix_T_piN}).
The corresponding results along the $\mu_N$ axis for two different
temperatures are presented in the upper panel of Fig.~(\ref{etamix_mu_piN}).
Lower panels of Fig.~(\ref{etamix_T_piN}) and middle panel of 
Fig.~(\ref{etamix_mu_piN}) are displaying the viscosity to entropy density
ratios as a function of $T$ (at three different values of $\mu_N$) and 
$\mu_N$ (at two different values of $T$). One should comparatively notice
the dash-dotted line in the lower panel of Fig.~(\ref{eta_s_mu}) and the middle
panel of Fig.~(\ref{etamix_mu_piN}), which are exhibiting an opposite 
nature in the low $\mu_N$ region. Hence the  approximated $T$-$\mu_N$
range, where viscosity to entropy density ratio reduces, is transformed
from ($T=0-0.12$ GeV, $\mu_N=0-0.5$ GeV) to ($T=0-0.15$ GeV, $\mu_N=0-0.5$ GeV)
due to the mixing effect.
Being closer to the earlier results~\cite{Itakura,Denicol,Bass}, specially
the result of Itakura et al.~\cite{Itakura}, the mixing effect appears to be very
important. Though the ratios in both cases, without and with mixing effect
increase beyond the $\mu_N\approx0.5$ GeV but the conclusion of 
our results, based on the effective hadronic
Lagrangian, should be concentrated within regions 
of $0.100$ GeV $<T<0.160$ GeV and $0<\mu_N<0.500$ GeV.
The lower panel of Fig.~(\ref{etamix_mu_piN}) shows
the $T$-$\mu_N$ points where the ratios are approximately
equal to its KSS bound. From this plot we can get
a rough idea of $T$-$\mu_N$ region, where our shear
viscosity calculations for the pion-nucleon system
may be considered as a reliable estimation by using
the effective hadronic model.
%
\section{Summary and Conclusion}
\label{sec:concl}
The present work is an extension of our previous studies~\cite{GKS,G_N} of
the shear viscosity calculations for pionic~\cite{GKS} and nucleonic~\cite{G_N}
components, where pion thermal width due to different mesonic fluctuations and the nucleon
thermal width due to different pion-baryon fluctuations are respectively
considered. However, in the two component pion-nucleon system, pion thermal
width may also be originated from different baryonic loops, which
is not taken in our previous investigations~\cite{GKS,G_N}. 
Considering this baryonic loop contribution, we have addressed
a complete picture of pion and nucleon propagation via
all possible meson and baryon quantum fluctuations at finite
temperature and density, from where their corresponding
contributions to the shear viscosities have been found.

Following the traditional technique of Kubo relation~\cite{Nicola,Weise,S_rev,G_IJMPA}, 
the shear viscosities of pion and nucleon components can be deduced from
their corresponding correlators of viscous stress tensor in the static limit,
which will be non-divergent when a
finite thermal width will be introduced in their free propagators.
These finite values of pion and nucleon thermal widths have been
estimated from the RTF calculations of the pion self-energy for different mesonic and baryonic
loops and the nucleon self-energy for different pion-baryon loops.
Thermal width and its inverse quantity, mean free path for each component
is numerically generated as a function of the momentum $\vk$ of the constituent
and the medium parameters, $T$ and $\mu_N$. They show very non-trivial 
momentum distributions, which have been integrated out by the Bose-enhanced
and Pauli-blocked phase space factors of pion and nucleon, respectively, to calculate
their corresponding shear viscosities. We have plotted the shear viscosity
of each component and their total as a function of $T$ and $\mu_N$, where
one can observe a distinct and important effect of 
pion thermal width for baryonic loops, which is the main finding
of the present investigation to demonstrate a complete
picture of shear viscosity calculation for pion-nucleon system.
Actually the $\mu_N$ dependence is entering in the shear viscosity
of pionic component via this baryon loop contribution of pion thermal
width. This additional contribution makes the shear viscosity of pionic
component reduce with $\mu_N$ and increase with $T$.
By adopting a rough mixing effect of pion and nucleon densities
between two components, we have tried to present a numerical estimation of total 
shear viscosity for a mixed gas of pion-nucleon constituents.
Normalizing by the ideal expressions of entropy densities for pion
and nucleon gas, we have obtained the viscosity to entropy density ratios for each
component and their total. In the relevant $T$-$\mu_N$
region of hadronic domain, this ratio for
the pion-nucleon gas mixture reduces and approaches toward
its KSS bound as $T$ or $\mu_N$ increases.
%
%

{\bf Acknowledgment :}
This work is financed by Funda\c{c}\~ao de Amparo \`a Pesquisa do Estado de 
S\~ao Paulo - FAPESP, Grant Nos. 2012/16766-0.
I am very grateful to Prof. Gastao Krein for his academic
and non-academic support during my postdoctoral period in Brazil.
%
%
%

\end{document}